\documentstyle[twocolumn,aps,epsfig]{revtex}

\begin{document}
\title{A texture tensor to quantify deformations}
\author{Miguel Aubouy$^{(1)}$, Yi Jiang$^{(2)}$, James A.
Glazier$%
^{(3)}$, Fran\c{c}ois Graner$^{(4)}$}
\address{(1) SI3M\thanks{%
U.M.R. n${{}^{\circ }}$ 5819: CEA, CNRS and Univ. Joseph-Fourier.
Address for correspondence: maubouy@cea.fr.
Fax:
(+33) 4 38 78 56 91.
},
D.R.F.M.C., CEA, 38054 Grenoble Cedex 9, France\\
(2) Theoretical Division, Los Alamos National Laboratory, New Mexico 87545,
USA\\
(3)
Dept. of Physics, 316 Nieuwland, University of Notre Dame, Notre-Dame, IN
46556-5670, USA\thanks{
Present address: Department of Physics,
Swain West 159,
727 East Third Street,
Indiana University,
Bloomington, IN 47405-7105,
USA
}\\
(4) Laboratoire de Spectrom\'{e}trie Physique\thanks{%
U.M.R. n${{}^{\circ }}$ 5588 : CNRS and Univ. Joseph-Fourier.}, BP 87, 38402
St Martin d'H\`{e}res Cedex, France
\thanks{\noindent {\bf Acknowledgments: } We thank  M. Asipauskas, S.
Courty,  B.
Dollet, F. Elias, S. Ifadir,
E. Janiaud and G. Porte for discussions.  YJ is supported by the US DOE under
contract
W-7405-ENG-36. JAG acknowledges support from NSF, DOE  and NASA, the
French Minist\`{e}re de
l'Enseignement Sup\'{e}rieur, and hospitality at the LSP.
}}
%\date{\today}
\maketitle

\begin{abstract}

%{\bf Abstract: }
Under mechanical deformation, most materials exhibit both elastic and fluid
(or plastic) responses.
No existing formalism derived from microscopic
principles encompasses both their fluid-like and solid-like aspects.
We
define the {\it statistical texture tensor} to quantify the intuitive notion
of stored deformation. This tensor links microscopic and macroscopic
descriptions of the material, and extends the definition
of elastic strain.
\end{abstract}

%\vspace{0.4cm}

%\vspace{0.4cm}

%\narrowtext

%\newpage

A typical mechanical experiment applies a given macroscopic distortion to a
test sample and measures the resulting macroscopic force exerted by the
material, or {\it vice versa}. The goal is to find the constitutive
equation, which relates the macroscopic stress tensor to an independent
descriptor of the material's response. We know the relevant descriptor to
use for two extreme cases, elastic solids and isotropic fluids: the gradient
of the displacement and the velocity field, respectively.

While stress is  unambiguously defined
\cite{truesdell,alexander},  strain admits more than one definition.
Classical  linear (or even non-linear)  elasticity operationally 
defines the strain
by
comparing the current microscopic state to a fixed microscopic reference
state \cite{elasticity1,elasticity2,nonlinear}.
But most materials, having both an internal structure which stores elastic
energy, and the flexibility
to allow  rearrangements,
lie between ideal fluids and purely elastic solids. Not only do we lack
their exact constitutive relations, we do not know what descriptors apply.

In this paper we propose an
operational definition of the deformation which we can measure in
experiments and
simulations in terms of averages of microscopic quantities: the {\it
statistical texture tensor}, a state function of the material.  Its
variations measure the elastic strain of an object under arbitrary
deformations, without requiring the microscopic details of a
reference state.

\begin{figure}[h]
\centerline{
\epsfxsize 6truecm
\epsfbox{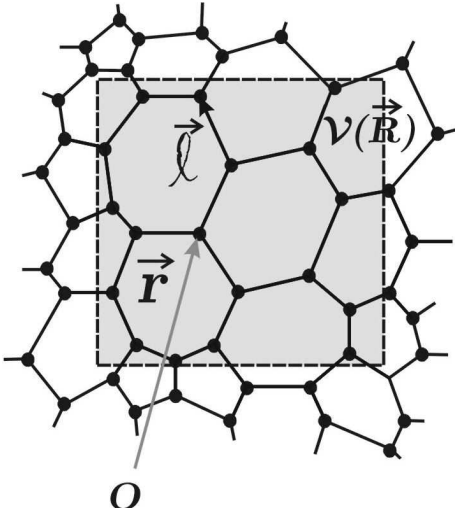}
}
\vspace{.5cm}
\caption{A network of interconnected sites. The representative volume
element  ${\cal V}$
at position $\vec{R}$ is a square in two dimensions, its volume is $V$.
Microscopically, $\vec{r}$ is the position of a site, and there are here
three links (unoriented vectors $\vec{\ell}$) per site. }
\label{defnetworkgeneral}
\end{figure}

Contrary to Ref. \cite{golden} which starts from a coarse grained mass density,
which is appropriate for granular materials, we consider a network structure.
Our generic material is a network of sites connected by links, which can
detach from and reattach to other sites (Fig. \ref{defnetworkgeneral}).
Site and link definitions depend on the material:

(i) In cellular
patterns, a site is the meeting point of cells. Two sites connect if their
cells share
an edge.
This concerns for instance grain boundaries
in crystals \cite{durand}, compact 2D or 3D aggregates of biological 
cells, or Voronoi tesselation.

(ii) If links are physical
objects, sites need not be. In liquid foams, links are bubble edges, 
while sites
are the vertices where they meet. In gels of polymers, links are
macromolecules; sites are knots.

(iii) If sites are undeformable objects, as in hard granular materials,
they link if their separation is less than a cut-off distance.
Since such cut-off is arbitrary, it must be chosen consistently throughout
the analysis.

(iv) When two sites
exert a force on each other, they are linked.
This is the case {\it e.g.} for atoms or molecules in
crystalline and amorphous solids, or deformable granular materials.
We then have to specify an cut-off   on their interaction
force  (Cauchy scheme): again,  such arbitrary cut-off must be chosen 
consistently.

Note that, even in cases (ii) and (iv), the
strain remains purely geometric and does not depend explicitly on stresses
and forces. This is seen in 3D foams where the films determine the stress
\cite{kraynik} while the edges determine the statistical strain.

For simplicity, we consider only point-like isotropic sites, {\it i.e.} the
average link length is much larger than the site size, and the unstressed
material is mesoscopically isotropic \cite{isoref}. Both restrictions are
inessential: we could relax them by extending our definitions.

At the microscopic level, we describe our material by the positions $\{\vec{r%
}_{s}\}$ of the sites $\{s\}$ in a $d $-dimensional space (Fig. \ref%
{defnetworkgeneral}). The {\it topology} ${\cal T}$ is the list of all
pairs $(s,s^{\prime })$ of sites connected by a link $\vec{\ell}=(\ell
_{x},\ell _{y},...)=\vec{r}_{s^{\prime }}-\vec{r}_{s}$. The topology changes
when perturbations create or destroy a link or  a site.

%\newpage

A continuous description interpolates between this microscopic level of
description and a macroscopic description in terms of the shape and size of
the network boundaries. We define an intermediate mesoscopic level (Fig.
\ref%
{defnetworkgeneral}). We cut the network into representative volume
elements, ${\cal V}({\vec{R}})$
of volume ${V}$, with positions $\vec{R}$: small enough that their
properties are constant over the box; but large enough that each piece
contains enough links to compute statistical properties and average out
microscopic details. In each volume element we define all
statistical quantities as averages over all links in ${\cal V}({\vec{R}})$.
The number of neighbours of each site must be much smaller than the number
of links in ${\cal V}({\vec{R}})$. We thus require that the topology remains
{\it ``short-range:''} while the neighbours $s^{\prime }$ of a site $s$ may
change, their number must remain bounded.

\begin{figure}[h]
\centerline{
\epsfysize  8truecm
%\epsfbox{ssave.eps}
%\rotatebox{-90}{\includegraphics{ssave.eps}}
\rotatebox{-90}{\epsfbox{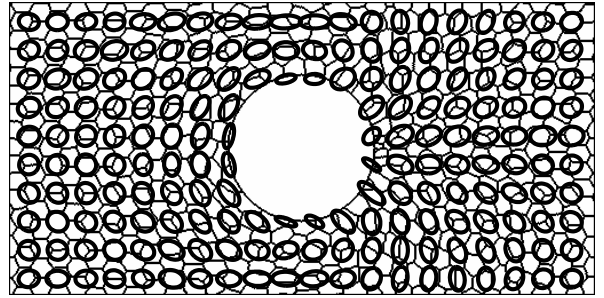}}
}
\vspace{.5cm}
\caption{Texture tensor $\stackrel{\_}{\bar{M}}$ (eq. \protect\ref{def2})
superimposed on a snapshot of a simulated
two-dimensional foam.
The foam flows steadily from left to right around a fixed round obstacle. We
calculate the texture tensor over a box almost as small as the bubble size,
by averaging over 50 successive images. The texture tensor is symmetric with
two positive eigenvalues. We represent it as an ellipse, with the long axis
(resp. small axis) proportional in length to the largest (resp. smallest)
eigenvalue and pointing along the corresponding eigenvector. We see
compression defomation in front of the obstacle, and stretching behind it. }
\label{flowobstacle}
\end{figure}

Our fundamental definition is the {\it texture tensor }:
\begin{equation}
\stackrel{\_}{\bar{M}}(\vec{R})\equiv \left\langle \;\vec{\ell}\otimes
\vec{%
\ell}\;\right\rangle .  \label{def2}
\end{equation}
Here $(\vec{\ell}\otimes \vec{\ell})_{ij}=\ell _{i}\ell _{j}$ is the
``dyadic'' (or ``outer'' or ``tensor'') product. $\stackrel{\_}{\bar{M}}$ is
symmetric, and has positive eigenvalues. $\stackrel{\_}{\bar{M}}$ occurs in
many different physical contexts: the {\it Steiner Tensor }
\cite{steiner,muller,duda},
order parameters of nematics \cite{nematics}, molecular moments of inertia
($%
\!$\cite{volino} p. 116-119), or textures of granular materials
\cite{texture}. The texture tensor quantifies our mental image of a network;
large eigenvalues correspond to directions of stretching (Fig.
\ref{flowobstacle}).

This definition (\ref{def2}) encompasses all materials where a mesoscopic scale
exists, {\it i.e.}, most cases. It requires that a thermodynamic limit
exists for all extensive and intensive quantities, {\it i.e.} statistics on
larger volumes decrease the relative amplitude of fluctuations (these
microscopic fluctuations might remain visible on the scale of the sample
size if the sample is small \cite{Pottsdyn} or if the lattice is ordered
\cite{khan}). We assume no correlations between
the volume elements, which we must check. Materials that display
networks of forces, avalanches or fractures thus require careful treatment.
For materials in ergodic steady flow, time averages also reduce fluctuations
and the network acts as a continuous medium down to scales as small as the
average link length (Fig. \ref{flowobstacle}).

If each region has an isotropic reference state \cite{isoref}, then
$\stackrel{\_}{\bar{M}}_{0}=M_{0}\stackrel{\_}{\bar{I}}%
_d$,
where $\overline{\overline{I}}$ is the identity tensor  in $d$ dimensions.
All its
relevant information
  lies in the scalar $M_{0}=\left\langle \ell_{0}^{2}\right\rangle/d $,
{\it i.e.} in the mean squared link length in this state \cite{optical}.

We now define the statistical strain $\stackrel{\_}{\bar{U}}$ as:
\begin{equation}
\stackrel{\_}{\bar{U}}(\vec{R})\equiv \frac{\log \stackrel{\_}{\bar{M}} -
\log \stackrel{\_}{\bar{M}}_{0}}{2},  \label{defUiso}
\end{equation}
where the tensor $\log \stackrel{\_}{\bar{M}}$ has the same axes as $%
\stackrel{\_}{\bar{M}}$ but eigenvalues equal to the logarithm of those of $%
\stackrel{\_}{\bar{M}}$.

Eq. (\ref{defUiso}) is operational: $\stackrel{\_}{\bar{U}}$ quantifies the
deformations visible locally in each region of an image such as extension,
compression, shear, or dilation (Fig. \ref{flowobstacle}). $\stackrel{\_}{%
\bar{U}}$ also satisfies other requirements for the definition of 
strain. It is invariant under
rotation, translation and transposition of indices. It is a true function of
state: it depends explicitly only on the current state and the reference
state, not on the detailed history of the material. It is mesoscopic: it
reflects only statistically significant features of the network. Its 
definition (eq.
\ref{defUiso}) accomodates topological changes and thus the plastic regime.

Finally, we prove
that
whenever the classical definition of the strain tensor applies, it coincides
with our statistical strain $\stackrel{\_}{\bar{U}} $.
Under an arbitrary elastic deformation, from an initial, equilibrium,
isotropic state  to a final state,
we have:
\begin{equation}
\stackrel{\_}{\bar{U}}=\frac{1}{2}\left( \stackrel{\_}{\bar{w}}+\stackrel{\_%
}{\bar{w}}^{t}\right) +O\left( w^{2}\right), \label{dU,w}
\end{equation}
   Here  we do not
make any assumption about the microscopic displacements.
  The displacement field $\vec{u}(\vec{R})$ is the difference between
states of the region ${\cal V}({\vec{R}})$ initially at position $\vec{R}$.
When $\vec{u}(\vec{R})$ is differentiable, the distortion tensor $%
\stackrel{\_}{\bar{w}}=[\vec{\nabla}\otimes \vec{u}]^{t}= \partial u_i /
\partial r_j$ quantifies the deformation from initial to final state.
The proof goes as follows.

As a first step,  we have to prove that the average over the surface 
element $\stackrel{\rightarrow
}{n}%
dS$ (or line elements, if in two dimension), denoted $\left\langle
.\right\rangle ^{\stackrel{\rightarrow }{n}}$, of
any function $g(\vec{\ell})$ of the links which actually cross this surface,
relates to its bulk average,
denoted $\left\langle .\right\rangle $, as:
\begin{equation}
\forall \stackrel{\rightarrow }{n},\quad \left\langle g\right\rangle ^{%
\stackrel{\rightarrow }{n}}\left\langle \ell _{j} \right\rangle
n_{j}=\left\langle g\ell _{j} \right\rangle n_{j}.  \label{demo}
\end{equation}
Here $\left\langle  \ell _{j} \right\rangle $ depends on the orientation
chosen for the link: it must be chosen consistently along the surface, for
instance along the orientation of $\stackrel{\rightarrow
}{n}$. On the opposite, if $g$ is an odd function of the link $\vec{%
\ell}$, {\it i.e.} $g(-\vec{\ell})=-g(\vec{\ell})$, then
   the quantity $g(\vec{%
\ell})\vec{\ell}$ does not depend on the arbitrary orientation choosen for
the link; in that case, $\left\langle g\ell _{j} \right\rangle $ does 
not depend
on $\stackrel{%
\rightarrow }{n}$, so that we can omit $\stackrel{%
\rightarrow }{n}$ in what follows.

To prove this first step, eq. (\ref{demo}), we remark that for any 
elementary section $\stackrel{%
\rightarrow }{dS}=\stackrel{\rightarrow }{n}dS$ of ${\cal C}$, oriented by
the unit vector $\stackrel{\rightarrow }{n}$, the probability for a link to
cross $\stackrel{\rightarrow }{dS}$ is proportional to: $$\rho \vec{\ell}
\cdot
\stackrel{\rightarrow }{n}\,dS\,P(\vec{\ell})\,d^{3}\vec{\ell},$$ where
$\rho
$ is the average density of links, and $P$ is the probability distribution
function of the links $\vec{\ell}$. Hence the sum of $g$ taken over all
links crossing $\stackrel{\rightarrow }{dS}$ is: $\sum_{\stackrel{%
\rightarrow }{dS}} g(\vec{\ell})= \rho \int \! \! \int \! \! \int \;g(\vec{%
\ell})\,\vec{\ell} \cdot \stackrel{\rightarrow }{n} \,dS\,P(\vec{\ell})\,d^{3}%
\vec{\ell}. %\label{sumdS}
$ In particular, with $g=1$ we obtain the number of links crossing $%
\stackrel{\rightarrow }{dS}$: $$\rho \int \! \! \int \! \! \int %\iiint
\;\vec{\ell}.\stackrel{\rightarrow}{n} \,dS\,P(\vec{\ell})\,d^{3}\vec{\ell}.$$
Combining both proves eq. (\ref{demo}).

At this point, it is interesting to apply eq. (\ref{demo}) to the 
forces. By definition \cite{elasticity1,elasticity2}, the stress
$\stackrel{\_}{%
\,\bar{\sigma}}$ is the tensor such that, $\forall \stackrel{\rightarrow }{n}
$:
  $$\sigma _{ij}n_{j}\,dS=\sum_{\stackrel{\rightarrow }{dS}}\tau_{i},$$ where
$\stackrel{\rightarrow }{\tau}$ is the tension (force) supported by the link
$\vec{\ell}$. With $g(\vec{\ell})=\tau_{i}$, eq. (\ref{demo}) yields:
%\begin{equation}
$$\stackrel{\_}{\,\bar{\sigma}}=\rho \langle \stackrel{\rightarrow
}{\tau}%
\otimes \vec{\ell} \rangle, $$ %\label{stress}
%\end{equation}
providing a statistical interpretation of the stress tensor as an average
over all links\cite{kraynik}.

Now, we can achieve the proof of eq. (\ref{dU,w}).
  By definition of $\vec{u}$ and $%
\stackrel{\_}{\bar{w}}$, we can write:
\begin{eqnarray}
\left\langle \delta \vec{\ell}\right\rangle ^{\stackrel{\rightarrow }{n}}
&=& \left\langle \delta \! \stackrel{\rightarrow }{r}\right\rangle
^{\stackrel{%
\rightarrow }{n}}(\stackrel{\rightarrow }{R}+ \langle \vec{\ell} \rangle ^{%
\stackrel{\rightarrow }{n}})-\left\langle \delta \! \stackrel{\rightarrow }{r}%
\right\rangle ^{\stackrel{\rightarrow }{n}}(\stackrel{\rightarrow }{R})
\nonumber \\
&=& \vec{u}(\stackrel{\rightarrow }{R}+ \langle \vec{\ell} \rangle ^{%
\stackrel{\rightarrow }{n}})-\vec{u}(\stackrel{\rightarrow }{R})  \nonumber
\\
&=& \stackrel{\_}{\bar{w}}^{t}\left\langle \vec{\ell}\right\rangle ^{%
\stackrel{\rightarrow }{n}} +O\left( w^{2}\right).  \label{subaffinity}
\end{eqnarray}
With $g(\vec{\ell})=\delta \ell _{i}$, eq. (\ref{demo}) yields: $$%
\left\langle \delta \vec{\ell}\otimes \vec{\ell}\right\rangle =\left\langle
\delta \vec{\ell}\right\rangle ^{\stackrel{\rightarrow }{n}}\otimes
\left\langle \vec{\ell}\right\rangle, $$ which inserted in eqs. (\ref{def2})
and (\ref{subaffinity}) yields:
\begin{eqnarray}
\stackrel{\_}{\bar{M}}&= &
\stackrel{\_}{%
\bar{M}}_{0} +
\stackrel{\_}{%
\bar{M}}_{0}\stackrel{\_}{\bar{w}}+\stackrel{\_}{\bar{w}}^{t}\stackrel{\_}{
\bar{M}}_{0}
\nonumber \\
&=& M_{0}\left( \stackrel{\_}{\bar{I}}+\stackrel{\_}{\bar{w}}+%
\stackrel{\_}{\bar{w}}^{t}\right)+O\left( w^{2} \right).
\label{arrayM}
\end{eqnarray}
Combining eqs.   (\ref{defUiso}) and (\ref{arrayM})
  proves eq. (\ref{dU,w}).

Hence the statistical strain $\stackrel{\_}{\bar{U}}$   coincides 
with the elastic
strain $\stackrel{\_%
}{\bar{u}}$.
Note that in the (rare) cases where the microscopic displacements
are affine, then $\vec{\ell}=(\stackrel{\_}{\bar{I}}+\stackrel{\_}{\bar{w}}%
^{t})\; \vec{\ell}_{0}$, $\forall \vec{\ell}$. This is a very strong 
statement since
each link (and not only the average) obeys eq. (\ref{subaffinity}); 
in this case,
eq. (\ref{dU,w}) is easier to demonstrate.

In summary, we have proposed a statistical characterization of
deformation. It averages the microscopic details of the current pattern, and of
the reference state, to keep only the physical features statistically 
relevant at
large scale.  Hence different microscopic configurations which are 
statistically
identical correspond to the same statistical strain.
For instance, ductile metals like soft steel or
aluminium have almost unchanged Young's modulus even much beyond their
yield
strain \cite{massonet}; then, although different applied strains correspond to
different microscopic
structures, they have the same statistical strain, the same static stress
and the same mechanical response for any physically reasonable static
constitutive relation.
Or, in a steadily
flowing material, $\stackrel{\_}{\bar{U}}$ is constant; we thus provide an
operational definition for the thermodynamic stored strain that Porte{\it \
et al.} \cite{porte} introduced to allow a theoretical description of
shear-induced phase transitions.

This definition invites re-analysis of existing data, as well as
experimental, numerical and theoretical tests. We have sucessfully performed
such tests on an experiment which forces a two-dimensional foam to flow
through a small constriction in a companion paper \cite{preparation}.

%\newpage

%\begin{references}


\begin{references}
\bibitem{optical}  $\stackrel{\_}{\bar{M}}$ distinguishes the anisotropy of
rank two tensors (``optical anisotropy''). For instance, $\stackrel{\_}{\bar{%
M}}$ distinguishes between birefrigent and non-birefringent networks. In an
``optically'' isotropic state, $\stackrel{\_}{\bar{M}}$ has two identical
eigenvalues and reduces to a scalar:
$\stackrel{\_}{\bar{M}}(\vec{R})=M(\vec{%
R})\stackrel{\_}{\bar{I}}_{d}$, where $\stackrel{\_}{\bar{I}}_{d}$ is the
identity tensor in $d$ dimensions. Since the trace of $\stackrel{\_}{\bar{M}}
$ is $\left\langle \ell _{i}\ell _{i}\right\rangle =\left\langle \ell
^{2}\right\rangle $, then $M=\left\langle \ell ^{2}\right\rangle /d$. Note,
however, that $\stackrel{\_}{\bar{M}}$ does not distinguish the anisotropy of
rank four tensors (``mechanical anisotropy''): for instance, $\stackrel{\_}{%
\bar{M}}$ does not necessarily distinguish between crystals and glasses.
\end{references}
\end{document}